# Wall slip effects on the fiber orientation of a short-fiber suspension in hyperbolic channel flow

**Kostas D. Housiadas***
*Department of Mathematics, University of the Aegean, Karlovassi, Samos, 83200, Greece*
*\*Corresponding author, email: housiada@aegean.gr*

**Antony N. Beris**
*Center of Research in Soft Matter and Polymers (CRISP) &*
*Department of Chemical and Biomolecular Engineering,*
*University of Delaware, Newark, Delaware 19716, USA*

**Suresh G. Advani**
*Center of Composite Materials & Department of Mechanical Engineering*
*University of Delaware, Newark, Delaware 19716, USA*

**Abstract**

We investigate the effect of wall fluid slip on the orientation of non-Brownian, short, rigid, and high aspect ratio cylindrical fibers suspended in a Newtonian fluid in flow through a symmetric hyperbolic planar channel. The fiber orientation is described using a second-order tensor formulation that accounts for fiber–fiber interactions and employs a hybrid closure to approximate the fourth-order orientation tensor, while neglecting the extra-stress contribution of the fibers to the total stress tensor. Building on our previous work on the no-slip case (Housiadas, Beris & Advani, *J. Rheol.*, 2025), the analytical Newtonian velocity field that has been obtained via the extended lubrication theory is utilized (Sialmas & Housiadas, *Eur. J. Mech/B* Fluids, 2024). Corresponding to this velocity field, the magnitude of the rate-of-deformation decreases as the slip coefficient increases. The resulting equations for the orientation tensor are solved numerically using a fully implicit finite difference method. The results show that the fiber orientation gradually evolves from its initial pure-shear state at the inlet toward a more aligned configuration as the channel exit is approached. The region of higher fiber alignment, that is most pronounced at the midplane where the flow is purely extensional, extends further toward the walls as the slip increases.

## 1. Introduction

The processing of short-fiber suspensions and composites is central to many emerging industrial applications—ranging from injection and compression molding to additive manufacturing [Šeta *et al.* (2023); Barakat *et al.* (2025)]—but remains challenging to model accurately. Predicting the evolution of fiber orientation during flow is crucial, as it governs the mechanical properties of the final material [Advani (1994); Bibbo *et al.* (1985); Djalili-Moghaddam & Toll (1996); Fu & Lauke (1996); Hu *et al.* (2024); Simon *et al.* (2020); Wang & Smith (2020); Lambert & Baird (2017); Hambach & Volkmer (2017); Pibulchinda *et al.* (2022); Yan *et al.* (2023)]. Fiber orientation is affected by the flow kinematics, fiber–fiber interactions, temperature effects, and boundary conditions, while in more concentrated systems, feedback effects of the fibers on the flow can also become significant. Despite the abundance of theoretical and computational studies available in the literature, the prediction of orientation in complex, non-viscometric flows remains a formidable task [Tucker (2022)], further emphasizing the need for the development of new methods and theoretical advances.

Among such complex flows, those occurring in hyperbolic geometries—such as symmetric 2D planar channels or 3D axisymmetric cylindrical pipes—are of particular interest. They provide fundamental models for extensional-dominated flows encountered in processes like 3D printing and fiber-reinforced extrusion. These geometries exhibit a transition from a purely shear flow near the walls to a strongly extensional flow along the midplane [Housiadas & Beris (2023)] or axis of symmetry [Housiadas & Beris (2024a)]. This characteristic makes them suitable for controlling fiber alignment through geometrical design, as recently shown in short-fiber suspension flows [Housiadas *et al.* (2025)].

The theoretical framework for modeling fiber orientation is based on the orientation tensor formalism introduced by Advani & Tucker (1987). This approach replaces the probability distribution function of fiber directions with its moment tensors, leading to an infinite sequence of orientation tensors. This hierarchy provides a macroscopic measure of fiber alignment, with the lowest-order members—the second- and fourth-order tensors—being the most significant. The second-order orientation tensor is the most fundamental, whereas the fourth-order tensor must be approximated in terms of the second-order tensor through a closure relation. Among the various closures proposed, the hybrid closure—a convex combination of the linear [Hand (1962)] and quadratic [Doi (1981)] closures—offers a satisfactory compromise between accuracy and computational simplicity, capturing both the fully random and perfectly aligned limits. It should be noted, however, that the development of accurate and efficient closure models remains an active area of research [Férec & Ausias (2015); Tucker (2022); Corona *et al.* (2023); Karl *et al.* (2023)].

For semi-dilute suspensions under isothermal conditions, the fiber extra-stress contribution to the total stress may be neglected, allowing the orientation problem to be treated independently of the flow field [Tucker (1991); VerWeyst & Tucker (2002)]. Based on this assumption and the Advani & Tucker (1987) formulation, Housiadas *et al.* (2025) developed a general framework for solving the orientation field of short fibers in a Newtonian matrix fluid flowing through a hyperbolic planar channel with no-slip boundary conditions. In that study, the velocity field was obtained using the high-accuracy extended lubrication theory of Sialmas & Housiadas (2024). The orientation tensor equations were solved for a new set of orientation components which removes the jump discontinuities observed at the inlet with the original components, and ensures continuity along the channel walls. The simulations revealed that along the walls the fiber orientation remains constant (corresponding to pure shear flow) depending solely on the fiber–fiber interaction coefficient, whereas along the midplane it is continuously evolving downstream following the extensional nature of the flow. As shown by Sialmas & Housiadas (2024), the effective extensional region is restricted in a thin layer around the midplane and was found to shrink with increasing contraction ratio and to expand with decreasing aspect ratio of the channel. The fiber orientation state was also shown by Housiadas *et al.* (2025) that follows very closely that of the fluid deformation. These results demonstrated that the wall orientation profiles dominate the cross-sectional evolution of the fiber orientation, except near the midplane, where extensional alignment prevails. Overall, these findings provided critical insights into how geometrical parameters control fiber alignment in hyperbolic channels, thereby laying the foundation for the present investigation of wall-slip effects.

In the present work, we extend that analysis by incorporating wall slip along the channel walls, using the Newtonian kinematics previously derived by Sialmas & Housiadas (2024) based on Navier's slip law [Navier (1827)]. While the no-slip condition is generally valid for Newtonian fluids over untreated or rough surfaces, it may break down for polymeric and other complex fluids and/or suitably treated surfaces [Brenner & Ganesan (2000); Granick *et al.* (2003); Hatzikiriakos (2015)]. In such systems, wall slip is widely observed and manifests macroscopically as an effective slip velocity or slip length at the boundary, substantially modifying the local balance between shear and extension by reducing near-wall shear and influencing the flow throughout the domain [Ramamurthy (1986); Denn (2001); Hatzikiriakos (2012)]. In the limiting case of perfect slip, the flow structure is simplified due to a reduction in effective dimensionality, as demonstrated in applications such as compression molding of short-fiber composites [Barone & Caulk (1986)]. More generally, recent three-dimensional simulations of confined flows, including thin-wall injection molding of short-fiber composites, have shown that slip, when imposed as a boundary condition, can significantly alter the resulting flow field [Jiang *et al.* (2022)]. These observations motivate a systematic investigation of wall-slip effects in flows with mixed shear and extensional kinematics relevant to fiber-reinforced materials.

To focus on fundamental orientation trends while maintaining the hybrid analytical–numerical approach, three assumptions—previously also adopted by Housiadas *et al.* (2025)—are made: fibers are modelled in the infinite-aspect-ratio limit; the fiber–fiber interaction coefficient is constant and independent of flow type; and fiber extra-stresses are neglected in the momentum balance (decoupled approach). These simplifications allow us to isolate the influence of wall slip on fiber orientation without the additional complexity of fully coupled or fiber finite-aspect-ratio models. Justification for these assumptions is provided in the main text and supported by results presented in Appendix B.

The remainder of the paper is organized as follows. Section 2 presents the problem formulation, Section 3 outlines the solution methodology, Section 4 discusses the results, and Section 5 summarizes the main conclusions. Appendix A collects supplementary equations, while Appendix B provides a detailed justification of the fiber infinite–aspect–ratio approach adopted here.

## 2.  Problem formulation

We consider an incompressible dilute or semi-dilute suspension consisting of an incompressible Newtonian matrix fluid with mass density $\rho^*$ and shear viscosity $\eta_s^*$ (throughout the paper, a star superscript denotes a dimensional quantity) along with short, rigid, slender, axisymmetric (cylindrical) fibers. The fluid enters a symmetric planar channel under creeping conditions at a constant volumetric flow rate per unit length in the neutral direction, $Q^*$, while constant temperature is maintained in both the fluid and the channel.

The channel consists of an entrance region with two parallel walls at a constant distance $2h_0^*$ apart, and a varying region of length $\ell^*$, with height $2h_0^*$ at the inlet of the varying region and $2h_f^*$ at the outlet; see Fig. 1. Thus, the suspension enters the hyperbolic section of the channel smoothly and in a fully developed state, without vortical or secondary motion, as observed in the sudden contraction–expansion geometry [Lipscomb *et al.* (1988); VerWeyst & Tucker (2002)]. This ensures a smooth transition from the entrance region to the hyperbolic section, without any flow instabilities or vortex formation at the inlet. As the fluid progresses through the channel, the fibers are convected with the flow and continuously adjust their orientation in response to the acting forces and mutual interactions.

We use a Cartesian coordinate system $(z^*, y^*, x^*)$ to describe the flow field, where $z^*$, $y^*$, and $z^*$ are the main, vertical, and neutral directions, respectively (see Fig. 1), with unit vectors where $\mathbf{e}_z, \mathbf{e}_y$ and $\mathbf{e}_x$ are the corresponding unit vectors. The walls of the hyperbolic channel are described by the shape function

$$H^*(z^*) = h_0^* \big/ \left(1 + (\Lambda - 1)(z^*/\ell^*)\right)$$

where $0 \leq z^* \leq \ell^*$ and $\Lambda \equiv h_0^*/h_f^*$ is the contraction ratio (for $\Lambda > 1$); note that $\Lambda = 1$ corresponds to a straight channel, and $0 < \Lambda < 1$ to an expanding channel.

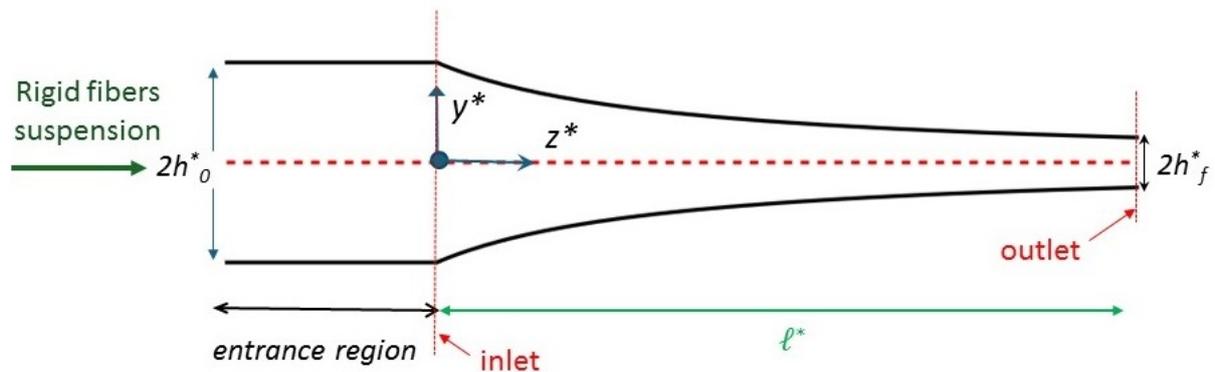

**Figure 1:** Geometry and Cartesian coordinate system (*y\*,z\**) for a symmetric hyperbolic channel.

Assuming two-dimensional flow and ignoring the variations of the field variables in the neutral direction, the velocity vector and total pressure are denoted by $\mathbf{u}^* = V^*(y^*,z^*)\mathbf{e}_y + U^*(y^*,z^*)\mathbf{e}_z$ and $p^* = p^*(y^*,z^*)$, respectively. Using the gradient operator $\nabla^* \equiv \mathbf{e}_y(\partial/\partial y^*) + \mathbf{e}_z(\partial/\partial z^*)$, the unit tensor $\boldsymbol{\delta}$, the rate of deformation tensor $\dot{\boldsymbol{\gamma}}^* = \nabla^*\mathbf{u}^* + (\nabla^*\mathbf{u}^*)^T$, and the associated extra-stress tensor $\boldsymbol{\tau}^*$ due to the fiber-flow interactions, we define the total stress tensor in the fluid as follows:

$$\boldsymbol{\sigma}^* := -p^*\boldsymbol{\delta} + \eta_s^*\dot{\boldsymbol{\gamma}}^* + \boldsymbol{\tau}^* \qquad (1)$$

where $\boldsymbol{\tau}^*$ is given by:

$$\boldsymbol{\tau}^* = \eta_s^* N_p \mathbf{A} : \dot{\boldsymbol{\gamma}}^* \qquad (2)$$

In the above equation, $N_p$ is a dimensionless fiber parameter related to the fiber volume fraction in the suspension and the fiber aspect ratio [Tucker (1991, 2022)]. In the absence of any external forces and torques, the mass and momentum balances for incompressible, steady-state and inertialess flow are:

$$\nabla^* \cdot \mathbf{u}^* = 0, \quad \nabla^* \cdot \boldsymbol{\sigma}^* = \mathbf{0} \qquad (3a,b)$$

The orientation of the fibers is described through the evolution equation of the second-order orientation tensor $\mathbf{a}$, using the Advani & Tucker (1987) model. This model, for cylindrical, infinitely slender axisymmetric rigid fibers is:

$$\frac{D\mathbf{a}}{Dt^*} - \mathbf{a} \cdot \nabla^*\mathbf{u}^* - (\nabla^*\mathbf{u}^*)^T \cdot \mathbf{a} + \mathbf{A} : \dot{\boldsymbol{\gamma}}^* = 2C_i \dot{\gamma}^* (\boldsymbol{\delta} - 3\mathbf{a}) \qquad (4)$$

where $D/Dt^* \equiv \mathbf{u}^* \cdot \nabla^*$ is the material derivative at steady state, and $\dot{\gamma}^* := \sqrt{\dot{\boldsymbol{\gamma}}^* : \dot{\boldsymbol{\gamma}}^*/2}$ is the magnitude of $\dot{\boldsymbol{\gamma}}^*$.

The right-hand side of Eq. (4) represents fiber–fiber interactions through an isotropic, phenomenological rotary diffusion term that drives the orientation distribution toward a random state. The diffusion coefficient $C_I \dot{\gamma}^*$ is taken in its simplest macroscopic form, assumed to scale linearly with the local flow intensity, quantified by the invariant magnitude of the rate-of-deformation tensor. The interaction coefficient $C_I$ is introduced as an effective phenomenological parameter controlling the strength of this diffusion, intended to capture the dominant orientation trends within a minimal constitutive framework. Several authors have also proposed expressions for $C_I$ in terms of fiber concentration and aspect ratio (see Fe'rec et al. (2009) for details and discussion). In the absence of fiber–fiber interactions ($C_I = 0$), Eq. (4) reduces to the Jeffery equation describing the kinematics of slender particles (corresponding to the fiber infinite–aspect–ratio limit) in a Newtonian fluid [Jeffery (1922)].

Although the flow in the hyperbolic section of the channel transitions from shear-dominated near the walls to purely extensional along the midplane, no explicit dependence of $C_I$ on flow type is introduced in the present model. This modeling choice is made to focus on the influence of wall slip on fiber orientation in flows with mixed kinematics, without introducing additional constitutive assumptions or requiring separate calibration of rotary diffusion in shear- and extension-dominated regions.

Equation (4) corresponds to the fiber infinite–aspect–ratio limit of the general orientation equation; both the finite- and infinite–aspect–ratio formulations are discussed in Housiadas *et al.* (2025). In applications such as additive manufacturing and 3D printing, fibers typically have large aspect ratios (of order twenty or larger). For the flow geometry considered here, our numerical results indicate that the influence of finite aspect ratio is small for values around ten and becomes negligible for values around twenty, for which the predicted orientation fields are indistinguishable, within numerical accuracy, from the infinite–aspect–ratio limit (see Appendix B). On this basis, the fiber infinite–aspect–ratio formulation of Eq. (4) is adopted in this work.

For the closure term, [Advani & Tucker (1987)] concluded that the quadratic closure [Doi (1981)] gives exact results for perfectly aligned fibers, while the linear closure [Hand (1962)] gives exact results for perfectly random fibers. Thus, the hybrid model, defined as the convex combination of those two [Advani & Tucker (1990); Tucker (2022)] appears to work better than both:

$$\mathbf{A} = (1-f)\mathbf{A}^{(L)} + f\,\mathbf{A}^{(Q)}, \tag{5}$$

where $\mathbf{A}^{(L)}$ and $\mathbf{A}^{(Q)}$ are the linear and quadratic parts, respectively. Denoting with $\det(\mathbf{a})$ the determinant of $\mathbf{a}$, $f$ is given as:

$$f = 1 - 27\det(\mathbf{a}) \tag{6}$$

The quadratic term $\mathbf{A}^{(Q)}$ is given as:

$$\mathbf{A}^{(Q)} = \mathbf{a}\mathbf{a} \tag{7}$$

The linear term $\mathbf{A}^{(L)}$ is given component-wise [Advani & Tucker (1990); Tucker (2022)]:

$$A^{(L)}_{ijkl} = -\frac{1}{35}\left(\delta_{ij}\delta_{kl} + \delta_{ik}\delta_{jl} + \delta_{il}\delta_{jk}\right) + \frac{1}{7}\left(a_{ij}\delta_{kl} + a_{ik}\delta_{jl} + a_{il}\delta_{jk} + a_{jl}\delta_{ik} + a_{jk}\delta_{il} + a_{kl}\delta_{ij}\right) \tag{8}$$

Therefore, the *ij*-component (where $i,j = x,y,z$) of the closure model in Eq. (5) is concisely written as:

$$(\mathbf{A}:\dot{\boldsymbol{\gamma}}^*)_{ij} = (1-f)(\mathbf{A}^{(L)}:\dot{\boldsymbol{\gamma}}^*)_{ij} + \mathbf{a}:\dot{\boldsymbol{\gamma}}^*\,f\,a_{ij} \tag{9}$$

The linear part, after performing the double dot product, gives:

$$(\mathbf{A}^{(L)}:\dot{\boldsymbol{\gamma}}^*)_{ij} = -\frac{2}{35}\dot{\gamma}^*_{ij} + \frac{1}{7}\left(2a_{ik}\dot{\gamma}^*_{jk} + 2a_{jl}\dot{\gamma}^*_{il} + a_{kl}\dot{\gamma}^*_{kl}\delta_{ij}\right) \tag{10}$$

The domain of definition of Eqs. (1)-(10) is

$$\Omega^* = \{(z^*, y^*) \,|\, 0 < z^* < \ell^*, -H^*(z^*) < y^* < H^*(z^*)\}$$

while appropriate boundary conditions along the channel walls accompany the differential equations. Specifically, a no-penetration condition is imposed in the direction normal to the rigid walls, together with the standard Navier slip condition for the tangential velocity component:

$$\mathbf{N}_{\pm} \cdot \mathbf{u}^* = 0, \quad \mathbf{T}_{\pm} \cdot \mathbf{u}^* = k^* (\boldsymbol{\sigma}^* \cdot \mathbf{N}_{\pm}) \cdot \mathbf{T}_{\pm} \tag{11}$$

where $\mathbf{N}_{\pm}$ and $\mathbf{T}_{\pm}$ are the unit normal and tangential vectors, respectively, pointing away from the fluid at the upper ($y^* = H^*(z^*)$) and lower ($y^* = -H^*(z^*)$) walls, and $k^*$ is the dimensional slip coefficient. Note that the subscript '+' denotes value at the upper wall, whereas the subscript '-' denotes value at the lower wall. Also, the integral constraint of mass due to fluid's incompressibility is utilized, $Q^* = \int_{\mathbf{S}^*_{in}} \mathbf{u}^* \cdot d\mathbf{S}^*$ =constant at any cross section of the channel, and a datum pressure, $p^*_{ref}$, is chosen at the wall of the outlet cross-section, i.e., $p^*_{ref} = P^*(H^*(\ell^*), \ell^*)$. Eq. (11) and the integral constraint hold for $0 \leq z^* \leq \ell^*$.

Finally, only initial conditions (in a Lagrangian sense) are required for the orientation tensor at the inlet of the varying section of the channel. These are specified based on the fully developed Poiseuille flow in the entrance region, as presented below.

## 2.1. Non-dimensionalization

Dimensionless variables are introduced by scaling $z^*$ by $\ell^*$, $y^*$ and $H^*$ by $h_0^*$, $U^*$ by $u_c^*$, $V^*$ by $\varepsilon u_c^*$, the pressure difference $p^* - p^*_{ref}$ by $\eta_s^* u_c^* \ell^* / h_0^{*2}$, where $u_c^* \equiv Q^* / h_0^*$, while the components of the orientation tensor are already dimensionless. Using these characteristic scales, the non-trivial components of the governing equations are derived and can be found component-wise in Housiadas *et al*. (2025). Note that in the dimensionless equations, the aspect ratio, $\varepsilon \equiv h_0^* / \ell^*$, of the varying region of the channel, appears. The aspect ratio is typically a small number ($\varepsilon \approx 0.1$ or less), which can be exploited to derive a simplified set of governing equations. This simplification, however, is applied only to the momentum balance, as explained below, in Section 3.

Based on the characteristic scales, the shape function $H$ becomes:

$$H(z) = \frac{1}{1+(\Lambda-1)z}, \quad 0 \leq z \leq 1$$

This shape function is continuous throughout the entire domain and infinitely differentiable in $z \in (0,1)$ but not differentiable at $z=0$ and $z=1$ because its slope exhibits a jump discontinuity. An interesting property of the hyperbolic geometry is that its derivative $H'(z)$ is expressed solely in terms of $H(z)$, i.e. $H'(z) = -(\Lambda-1)H^2(z)$. Note that $z=0$ marks the intersection between the entrance and varying regions of the channel, and $z=1$ corresponds to the outlet of the hyperbolic section.

The domain of definition of the governing equations is $\Omega = \{(z,y) | 0 < z < 1, -H(z) < y < H(z)\}$. The dimensionless no-penetration and slip conditions at the walls (at $y = \pm H(z)$) are $\mathbf{N}_\pm \cdot \mathbf{u} = 0$ and $\mathbf{T}_\pm \cdot \mathbf{u} = K(\boldsymbol{\sigma} \cdot \mathbf{N}_\pm) \cdot \mathbf{T}_\pm$, respectively, where $K := k^* \eta_s^* / h_0^*$ is the dimensionless slip coefficient. The integral constraint due to fluid's incompressibility becomes $\int_{-H(z)}^{H(z)} U(y,z) dy = 1$, $0 \leq z \leq 1$, and the reference pressure becomes $p(H(1),1) = 0$. Boundary conditions for the orientation tensor at the inlet, i.e., initial conditions in a Lagrangian sense, are derived from the analytical solution in the entrance region—the latter is provided in Section 4.

Typical values for the dimensional slip coefficient *K* are reported by Hatzikiriakos (2012; 1987) and Denn (2001), in the range $k^* = 10^{-5} - 10^{-3}$ m/(Pa s). Considering a typical Newtonian fluid (whose viscosity is usually close to $1\,\mathrm{Pa\,s}$) and a gap width of the channel a few millimeters, the dimensionless slip coefficient varies approximately in the range $0.001 < K < 0.5$.

### 3. Solution procedure

For the solution of the dimensionless governing equations, we use the methodology developed in previous works for the flow of Newtonian [29] and viscoelastic [21-23] fluids in hyperbolic geometries. It has been fully described by Housiadas *et al.* (2025) and is briefly presented here for completeness. First, we introduce the streamfunction, $\Psi$, defined with the aid of the two velocity components, i.e. $U = \partial \Psi / \partial y$ and $V = -\partial \Psi / \partial z$, so that the continuity equation is satisfied identically. Differentiating the *y*-component of the momentum balance with respect to *z*, the *z*-component with respect to *y*, and then subtracting the resulting equations yields a fourth-order partial differential equation for the streamfunction, in which the pressure has been eliminated:

$$\frac{\partial^2 \Psi}{\partial y^4} + 2\varepsilon^2 \frac{\partial^4 \Psi}{\partial y^2 \partial z^2} + \varepsilon^4 \frac{\partial^2 \Psi}{\partial z^4} + N_p \tau_f = 0, \qquad (12a)$$

$$\tau_f = \left( \frac{\partial^2}{\partial y^2} - \varepsilon^2 \frac{\partial^2}{\partial z^2} \right) \tau_{yz} + \varepsilon \frac{\partial^2}{\partial y \partial z} (\tau_{zz} - \tau_{yy}) \qquad (12b)$$

Second, we introduce new independent coordinates $(Y,Z)$ that map the varying boundaries of the flow domain into fixed ones, i.e., $Y = y/H(z)$ and $Z = z$. Using the new coordinates, the domain of definition of the governing equations becomes $\bar{\Omega} = \{(Z,Y) \mid 0 \leq Z \leq 1, -1 < Y < 1\}$. The first-order partial differential operators and the material derivative are transformed accordingly, in terms of the mapped coordinates:

$$\frac{\partial}{\partial y} = \frac{1}{H(Z)} \frac{\partial}{\partial Y}, \quad \frac{\partial}{\partial z} = \frac{\partial}{\partial Z} - \frac{Y H'(Z)}{H(Z)} \frac{\partial}{\partial Y}, \quad \frac{D}{Dt} = \frac{1}{H(Z)} \left( \frac{\partial \Psi}{\partial Y} \frac{\partial}{\partial Z} - \frac{\partial \Psi}{\partial Z} \frac{\partial}{\partial Y} \right) \qquad (13)$$

Obviously, for a straight channel ($H(z)=1$ or $\Lambda = 1$) the mapped coordinates and the differential operators are identical to the original ones; thus, the transformed equations are valid in both the entrance and varying sections of the channel.

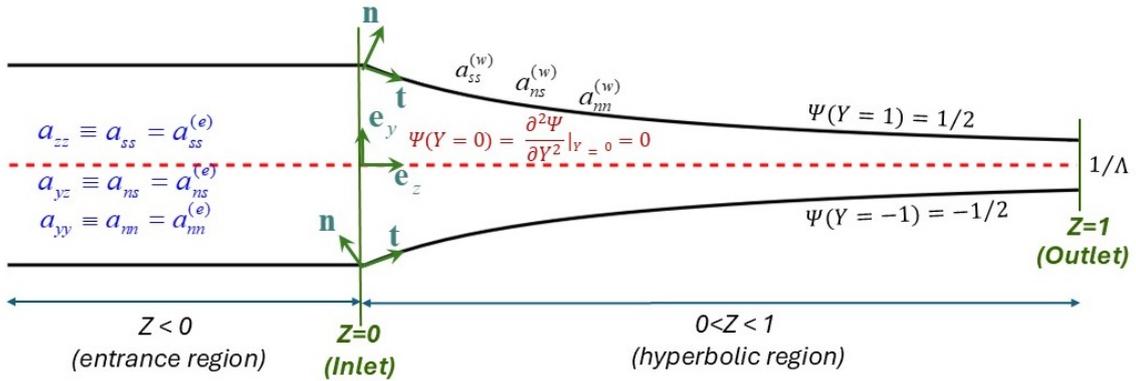

**Figure 2:** Conditions along the wall (*Y=1*) and on the midplane (*Y=0*) for the streamfunction *Ψ*. The orientation tensor in the entrance region (*Z<0*) is indicated by the superscript "*(e)*". For the straight region, the Cartesian and rotated components are identical. The Cartesian unit vectors ($\mathbf{e}_z, \mathbf{e}_y$) and the new set of orthogonal unit vectors ($\mathbf{t}, \mathbf{n}$) are also shown on the wall(s). The new unit vectors depend on both spatial coordinates *(Y,Z)*, while in the entrance region and on the midplane ($\mathbf{e}_z, \mathbf{e}_y$)=($\mathbf{t}, \mathbf{n}$).

Next, we consider the new set of unit vectors $\{\mathbf{t}, \mathbf{n}, \mathbf{e}\}$ (see Figure 2):

$$\mathbf{t} = \frac{\mathbf{e}_z - c\mathbf{e}_y}{\sqrt{1+c^2}}, \quad \mathbf{n} = \frac{c\mathbf{e}_z + \mathbf{e}_y}{\sqrt{1+c^2}}, \quad \mathbf{e} = \mathbf{e}_x \quad (14a)$$

$$c = c(Y,Z) = \varepsilon\,(\Lambda-1)Y H^2(Z) \quad (14b)$$

where the hyperbolic property $H'(Z) = -(\Lambda-1)H^2(Z)$ has been used to eliminate $H'(Z)$. For $Y$=constant where $-1 \le Y \le 1$, $\{\mathbf{t},\mathbf{n},\mathbf{e}\}$ form an orthonormal basis fitted on surfaces described by the original Cartesian coordinates as $y = Y H(z)$. At $Y = \pm 1$ these unit vectors are related to those on the upper and lower walls, i.e., $\mathbf{t}(Y=1) \equiv \mathbf{T}_+$, $\mathbf{n}(Y=1) \equiv \mathbf{N}_+$ and $\mathbf{t}(Y=-1) \equiv \mathbf{T}_-$, $\mathbf{n}(Y=-1) \equiv -\mathbf{N}_-$. Additionally, at $Y=0$ Eq. (14b) gives $c=0$, and hence $\{\mathbf{t},\mathbf{n},\mathbf{e}\} \equiv \{\mathbf{e}_z,\mathbf{e}_y,\mathbf{e}_x\}$. The latter holds in the entrance region too.

The new set of orthogonal unit vectors are used to find the components of the orientation tensor $a_{ss}$, $a_{ns}$, $a_{nn}$, and $a_{xx}$, which will hereafter be referred to as the *rotated components*, according to the expressions:

$$a_{ss} = (\mathbf{a}\cdot\mathbf{t})\cdot\mathbf{t}, \quad a_{ns} = (\mathbf{a}\cdot\mathbf{t})\cdot\mathbf{n}, \quad a_{nn} = (\mathbf{a}\cdot\mathbf{n})\cdot\mathbf{n}, \quad a_{xx} = (\mathbf{a}\cdot\mathbf{e})\cdot\mathbf{e} \quad (15)$$

Therefore, the orientation tensor $\mathbf{a}$, in terms of Cartesian components:

$$\mathbf{a} = \mathbf{e}_z\mathbf{e}_z a_{zz} + (\mathbf{e}_y\mathbf{e}_z + \mathbf{e}_z\mathbf{e}_y)a_{yz} + \mathbf{e}_y\mathbf{e}_y a_{yy} + \mathbf{e}_x\mathbf{e}_x a_{xx} \quad (16a)$$

is also expressed in terms of rotated components as:

$$\mathbf{a} = \mathbf{t}\mathbf{t}\,a_{ss} + (\mathbf{t}\mathbf{n}+\mathbf{n}\mathbf{t})a_{ns} + \mathbf{n}\mathbf{n}\,a_{nn} + \mathbf{e}\mathbf{e}\,a_{xx} \quad (16b)$$

Noting that $a_{xx}$ as defined in Eq. (15) is identical to the original xx-component of the orientation tensor, we use Eq. (15) to find $\{a_{zz}, a_{yz}, a_{yy}\}$ in terms of $\{a_{ss}, a_{ns}, a_{nn}\}$. This is achieved with the aid of the transformation matrix $\mathbf{m}$, and its inverse $\mathbf{m}^{-1}$, as $\mathbf{m}$ is constructed by performing the dot products in Eq. (15), resulting in:

$$\begin{pmatrix} a_{ss} \\ a_{ns} \\ a_{nn} \end{pmatrix} = \mathbf{m} \cdot \begin{pmatrix} a_{zz} \\ a_{yz} \\ a_{yy} \end{pmatrix} \Leftrightarrow \begin{pmatrix} a_{zz} \\ a_{yz} \\ a_{yy} \end{pmatrix} = \mathbf{m}^{-1} \cdot \begin{pmatrix} a_{ss} \\ a_{ns} \\ a_{nn} \end{pmatrix} \quad (17a)$$

where

$$\mathbf{m} = \frac{1}{1+c^2}\begin{pmatrix} 1 & -2c & c^2 \\ c & 1-c^2 & -c \\ c^2 & 2c & 1 \end{pmatrix} \Leftrightarrow \mathbf{m}^{-1} = \frac{1}{1+c^2}\begin{pmatrix} 1 & 2c & c^2 \\ -c & 1-c^2 & c \\ c^2 & -2c & 1 \end{pmatrix} \quad (17b)$$

Here $c$ is given Eq. (14b). Note that the inverse of $\mathbf{m}$ always exists since $\det(\mathbf{m}) = 1$, and we emphasize that $\mathbf{m}$ is continuous for $Z<0$ (for which $\mathbf{m} = \boldsymbol{\delta}$) and $Z>0$; hence the transformation introduced in Eq. (17a,b) cannot be applied at $Z=0$.

Eq. (17a,b) is used in the dimensionless version of Eq. (4) to derive the final equations in terms of the rotated components:

$$\varepsilon \frac{Da_{ss}}{Dt} - 2a_{ss}S_s^{(ss)} - 2a_{ns}S_n^{(ss)} + (\mathbf{A}:\dot{\boldsymbol{\gamma}})_{ss} = 2C_I \dot{\gamma}(1-3a_{ss}) \tag{18}$$

$$\varepsilon \frac{Da_{ns}}{Dt} + a_{ss}S_s^{(ns)} - a_{nn}S_n^{(ns)} + (\mathbf{A}:\dot{\boldsymbol{\gamma}})_{ns} = -6C_I \dot{\gamma} a_{ns} \tag{19}$$

$$\varepsilon \frac{Da_{nn}}{Dt} + 2a_{ns}S_s^{(nn)} + 2a_{nn}S_n^{(nn)} + (\mathbf{A}:\dot{\boldsymbol{\gamma}})_{nn} = 2C_I \dot{\gamma}(1-3a_{nn}) \tag{20}$$

$$\varepsilon \frac{Da_{xx}}{Dt} + (\mathbf{A}:\dot{\boldsymbol{\gamma}})_{xx} = 2C_I \dot{\gamma}(1-3a_{xx}) \tag{21}$$

where the closure terms $(\mathbf{A}:\dot{\boldsymbol{\gamma}})_{ij}$, $i,j = n,s,x$, are given as follows:

$$\begin{aligned}(\mathbf{A}:\dot{\boldsymbol{\gamma}})_{ij} &= (1-f)(\mathbf{A}^{(L)}:\dot{\boldsymbol{\gamma}})_{ij} + \mathbf{a}:\dot{\boldsymbol{\gamma}} f a_{ij} \\ f &= 1 - 27 a_{xx}(a_{ss}a_{nn} - a_{ns}^2) \\ \mathbf{a}:\dot{\boldsymbol{\gamma}} &= a_{nn}\dot{\gamma}_{yy} + 2a_{ns}\dot{\gamma}_{yz} + a_{ss}\dot{\gamma}_{zz}\end{aligned} \tag{22}$$

The components of the rate of deformation tensor are:

$$\begin{aligned}\dot{\gamma}_{yy} &= -\frac{2\varepsilon}{H}\frac{\partial}{\partial Y}\left(\frac{\partial \Psi}{\partial Z} - Y(\Lambda-1)H\frac{\partial \Psi}{\partial Y}\right) \\ \dot{\gamma}_{yz} &= \frac{1}{H}\frac{\partial}{\partial Y}\left(\frac{1}{H}\frac{\partial \Psi}{\partial Y}\right) - \varepsilon^2 \frac{\partial}{\partial Z}\left(\frac{\partial \Psi}{\partial Z} - Y(\Lambda-1)H\frac{\partial \Psi}{\partial Y}\right) \\ \dot{\gamma}_{zz} &= 2\varepsilon \frac{\partial}{\partial Z}\left(\frac{1}{H}\frac{\partial \Psi}{\partial Y}\right)\end{aligned} \tag{23}$$

In Eqs. (18)-(21), the stretching terms $S_s^{(ss)}, S_n^{(ss)}, S_s^{(ns)}, S_n^{(ns)}, S_s^{(nn)}, S_n^{(nn)}$, and linear closure terms $(\mathbf{A}^{(L)}:\dot{\boldsymbol{\gamma}})_{ij}$ are given in Appendix A. Note that due to the constraint $\sum_{i=s,n,x} a_{ii} = 1$, Eq. (21) is not used to determine the evolution of $a_{xx}$ but only as an individual check of the correctness and accuracy of the solution(s) derived here.

As seen above, $\mathbf{a}$ is given both in terms of its original Cartesian components and the rotated components; however, its physical interpretation is better understood through its eigenstructure, namely through its principal directions and principal values, which correspond to its eigenvectors $\{\mathbf{v}_1, \mathbf{v}_2, \mathbf{v}_3\}$ and eigenvalues $\{\zeta_1, \zeta_2, \zeta_3\}$, respectively. By convention, the first principal direction is associated with the largest eigenvalue of the tensor, while the third principal direction is associated with the smallest eigenvalue. The first principal direction of $\mathbf{a}$ corresponds to the direction along which the most fibers are aligned, while the third principal direction corresponds to the one along which the fewest fibers are aligned. When the orientation tensor $\mathbf{a}$ is given in any coordinate system, its diagonal components can be interpreted as the fraction of fibers oriented along the corresponding coordinate axis, while the off-diagonal components indicate that a principal direction of orientation is not aligned with the coordinates axes [Advani (1996); Tucker (2022)]. For perfectly random fibers the orientation tensor becomes fully isotropic, i.e., $\mathbf{a} = (1/3)\left(\mathbf{v}_1\mathbf{v}_1 + \mathbf{v}_2\mathbf{v}_2 + \mathbf{v}_3\mathbf{v}_3\right)$, while for fibers fully aligned in the main flow direction, $\mathbf{a} = \mathbf{v}_1\mathbf{v}_1$.

We close this subsection by noting that the third step of the methodology presented above has been applied to the leading-order lubrication equations for the viscoelastic flow of an Oldroyd-B fluid in a hyperbolic planar channel by Sialmas & Housiadas (2025) and Housiadas (2025). In those cases, the transformation matrix $\mathbf{m}$ and its inverse $\mathbf{m}^{-1}$ differ from those in Eq. (17b), because they correspond to the lowest-order lubrication approximation and different characteristic scales used to nondimensionalize each component of the polymer extra-stress tensor have been utilized. In the present work, however, the orientation tensor components are already dimensionless, and the full orientation tensor equations are solved.

### 3.1. Analytical approach for the flow-field

As mentioned in the Introduction, the present work does not account for the extra stress induced by fiber–flow interactions; instead, fiber orientation is predicted solely from the kinematics of the suspending Newtonian fluid; see Eqs. (12a,b). This is achieved by setting the fiber parameter $N_p = 0$. As shown by Tucker (1991), using the theory of Dinh & Armstrong (1984) for rigid fiber suspensions with Newtonian matrix fluids (see Fig. 1 in Tucker (1991) and Fig. 6.6 in Tucker (2022)), there exists a well-defined dilute-to-semi-dilute regime for fiber suspensions with aspect ratios larger than ten—values that, as shown in Appendix B, already closely approach the fiber infinite–aspect–ratio limit—in which $N_p < 1$. In this regime, the contribution of the fiber extra stress to the momentum balance, Eq. (12a), is small, and its neglect constitutes a reasonable first-order approximation. While fiber extra stresses may become locally significant in regions of predominantly extensional kinematics, such regions are spatially confined in the present flow geometry and configuration, and for the parameter range considered here their contribution to the overall momentum balance remains limited.

Coupled flow–orientation simulations of Newtonian fiber suspensions by Lipscomb *et al.* (1988) in a sudden axisymmetric contraction, and by VerWeyst & Tucker (2002) in an axisymmetric contraction (for $N_p = 6$), an axisymmetric expansion (for $N_p = 6$), and a center-gated disk (for $N_p = 19.35$), showed minor deviations from uncoupled simulations in the flow and orientation fields only in localized regions of the domain, primarily associated with vortex formation near corners or near the center of the gated disk.

Additionally, experimental and numerical studies of short-fiber viscoelastic suspensions (Egelmeers et al. 2024a,b, 2025) indicate that fiber-orientation kinetics under simple shear or uniaxial extension are largely insensitive to fiber concentration up to the onset of the concentrated regime. These findings provide qualitative support for the assumption that fiber extra-stresses have a limited influence on orientation trends in the dilute-to-semi-dilute regime and, within this context, may be neglected as a first-order approximation.

Setting $N_p = 0$, Eq. (12a) can be solved independently to determine the pressure and velocity fields in advance. This task was carried out by Sialmas & Housiadas (2024), who studied the creeping flow of a Newtonian fluid in hyperbolic tubes, exploiting the small magnitude of the tube's aspect ratio, $\varepsilon$. Using a high-order perturbation expansion in terms of $\varepsilon^2$, they obtained analytical solutions for the field variables up to the 20th order. By examining the convergence properties of the series solution, they found that the most accurate results—requiring the fewest terms in the series—were achieved with the first three terms. For the planar geometry, this solution can be safely applied for $(\Lambda-1)\varepsilon < 1$. The corresponding streamfunction $\Psi$, which is an antisymmetric function with respect to the midplane, i.e., $\Psi(-Y,Z) = -\Psi(Y,Z)$, satisfies $\Psi(Y=1,Z)=1/2$, $\Psi(Y=0,Z) = \partial^2\Psi/\partial Y^2\big|_{Y=0} = 0$, whereas all its derivatives with respect to $Z$ at the walls and the midplane are zero. It also satisfies the slip condition along the walls. For the hyperbolic geometry studied here, the solution, as found by Sialmas & Housiadas (2024), can be expressed in terms of *Y* and *H*; i.e., its dependence on Z is only implicit through *H=H(Z)*, and is given by:

$$\Psi(Y,H) \approx \psi_0(Y,H) + Y(1-Y^2)C^2(H)\left\{\psi_2(Y,H) + C^2(H)\psi_4(Y,H)\right\} \qquad (24)$$

where $C(H) \equiv \varepsilon(\Lambda-1)H^2$, and the leading-order term is:

$$\psi_0 = \frac{3(H+2K)Y}{4(H+3K)} - \frac{HY^3}{4(H+3K)} \qquad (25)$$

Functions $\psi_2$ and $\psi_4$ are too long to be printed here and are given in the Appendix. For the no-slip and perfect-slip cases, $\Psi$ reduces, respectively to:

$$\Psi(Y,H) \approx \frac{Y}{4}(3-Y^2) + \frac{3}{20}Y(1-Y^2)^2 C^2(H) + \frac{3}{50}Y(1-Y^2)^2 C^4(H), \quad K \to 0 \qquad (26a)$$

$$\Psi(Y,H) \approx \frac{Y}{2} + \frac{1}{3}Y(1-Y^2)C^2(H) - \frac{1}{15}Y(1-Y^4)C^4(H), \quad K \to \infty \qquad (26b)$$

### 3.2. Numerical approach for the orientation tensor

The numerical solution of the final equations for the orientation tensor is based on a fully implicit scheme that employs finite difference methods [Housiadas *et al.* (2025)]. Specifically, the integration along the Z-direction is performed as an initial value problem using at the inlet of the hyperbolic section (*Z*=0), the solution for simple Poiseuille flow. First, the computational domain [0,1] is discretized according to the equidistant grid points $Z_n = n/N$, $n = 0,1,2,...,N$, where $N$ is the total number of grid points in the Z-direction; thus $Z_0 = 0$ and $Z_N = 1$. The integration starts with the fully implicit A(0)-stable, first-order accurate, backwards differentiation formula (BDF), to approximate the Z-derivatives of the components of the orientation tensor. This formula is used only at the first few Z-steps to build the required information needed for the more accurate BDF formula (which is A(0) stable too). The *Y*-derivatives of $a_{ij}$ (needed for the evaluation of $Da_{ij}/Dt$) are calculated using the second-order central finite difference scheme for the equidistant points $Y_m = -1 + 2m/M$ where $m = 1,2,...,M-1$ and $M$ is even. At the boundaries ($m=0$ and $m=M$) and the midplane ($m = 1 + M/2$) the *Y*-derivatives are eliminated from the equations because they are multiplied with $\partial \Psi / \partial Z$ which is zero along the walls and on the midplane.

Starting from *n=1*, a strongly non-linear set of 3(*M*+1) algebraic equations is formed with unknowns being the orientation tensor components at the grid points $(Z_n, Y_m)$, where $m = 0,1,2,...,M$. The solution is found in a fully implicit manner for all grid points in the *Y*-direction using a Newton iterative scheme; the latter typically converges quadratically within three iterations, with an absolute error convergent criterion of 10$^{-14}$. Once the solution of the equations has been found at the *n*-th cross section, the algorithm proceeds with the calculations on the adjacent *k+1*-th cross section, until the terminal *N*-th cross section is reached (i.e., for $Z = Z_N = 1$).

Using *N*=101 and *M*=40 or 50, the performance and accuracy of the code was found to be excellent. All simulations were performed for the rotated components of the orientation tensor. Non-physical results in the diagonal components of the orientation tensor—previously reported in some cases by Lipscomb *et al.* (1988) and VerWeyst & Tucker (2002)—was not observed, for the range of parameters used here. In all cases, the numerical results satisfied $0 < a_{ii} < 1$ for $i = s, n, x$, as they should. Also, the eigenvalues of **a** were strictly positive confirming the positive definiteness property of **a** [Tucker 2022; Winters *et al.* (2022)].

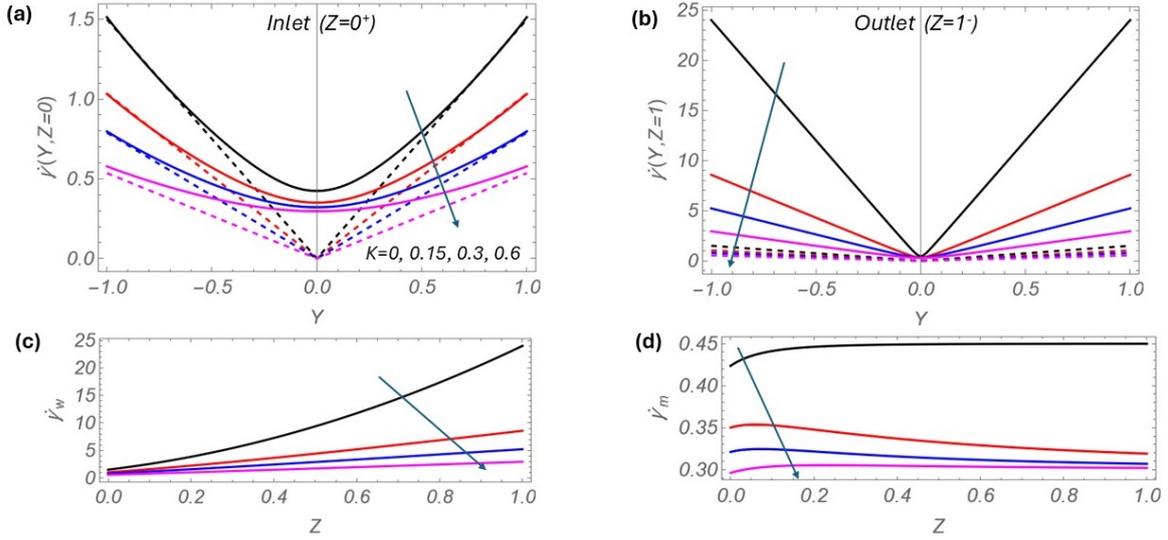

**Figure 3:** Effect of the slip coefficient K on the magnitude of the rate of deformation tensor for *ε*=0.1, $C_l$ = 0.05, and Λ=4.
(a) at the inlet (Z=0⁺) as a function of Y; and (b) at the outlet (Z=1⁻) as a function of Y.
Solid lines: hyperbolic section; Dashed lines: straight section.
(c) Along the wall(s) as a function of Z, (d) on the midplane as a function of Z.
Arrows indicate the direction of increasing *K* (*K* = 0, 0.15, 0.3, 0.6).

## 4. Results and discussion

### 4.1. Streamfunction and rate of deformation

First, we present results based on the Newtonian kinematics as derived by Sialmas & Housiadas (2024). At the entrance region, i.e., for $Z < 0$ (where *H*=Λ=1), the streamfunction $\Psi^{(e)}$ and the magnitude of the rate-of-deformation tensor $\dot{\gamma}^{(e)}$ are given by:

$$\Psi^{(e)}(Y) = \frac{3(1+2K)Y}{4(1+3K)} - \frac{Y^3}{4(1+3K)}, \quad \dot{\gamma}^{(e)}(Y) = \frac{3|Y|}{2(1+3K)} \tag{27}$$

where the superscript *"(e)"* is used hereafter to denote the entrance region; obviously, $\Psi^{(e)}$ is identical to $\psi_0$ given by Eq. (25) evaluated at $H = 1$.

The effect of the slip coefficient *K* on the magnitude of the rate of deformation tensor $\dot{\gamma}$, is presented in Figure 3 at the inlet (Z=0) and the outlet (Z=1) of the channel, as a function of the mapped vertical coordinate, *Y*. The discontinuity of $\dot{\gamma}$ before ($Z \to 0^-$) and after ($Z \to 0^+$) the inlet is clearly seen. For $Z \to 0^-$, $\dot{\gamma}$ varies linearly with *Y* (dashed lines), whereas for $Z \to 0^+$, a different profile is observed. As previously reported by Housiadas *et al.* (2025) for the no-slip case, this discontinuity is amplified with increasing contraction and/or aspect ratios, due both to changes in the *YZ* component and the additional components arising from the varying walls. Furthermore, increasing the slip coefficient reduces the magnitude of the rate-of-deformation tensor. This is expected, since fluid slip along the walls produces more uniform velocity profiles at any cross-section, resulting in smaller velocity gradients in both spatial directions. Thus, the difference between the midplane and wall values decrease as the slip coefficient increases. Interestingly, at the outlet of the hyperbolic section of the channel (Fig. 3b), the flow characteristics essentially resemble those of a simple Poiseuille flow, exhibiting a linear dependence on the *Y*-coordinate.

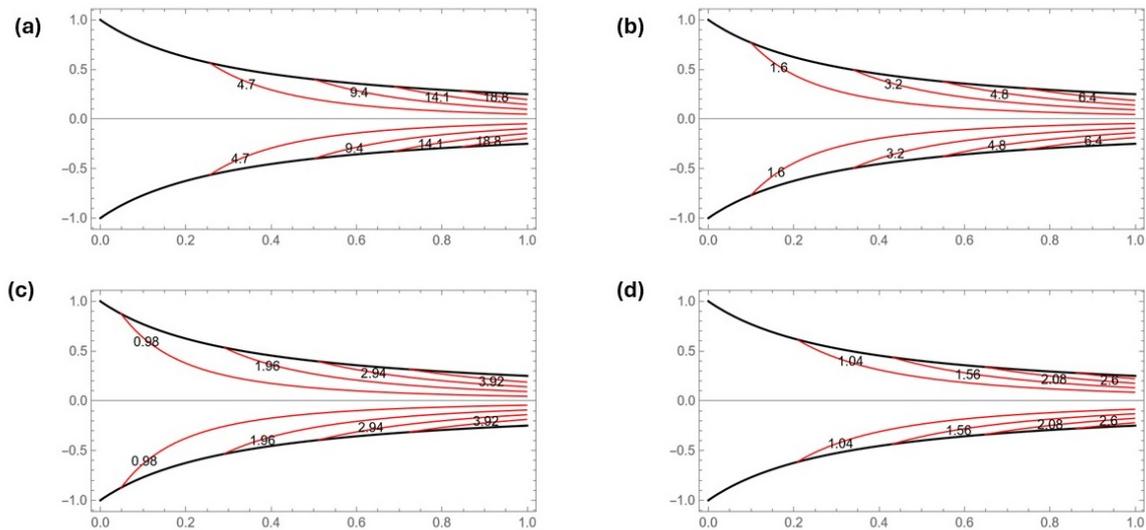

**Figure 4:** Contour plots of the magnitude of the rate of deformation tensor in the hyperbolic section. (a) *K* = 0; (b) *K* = 0.15; (c) *K* = 0.3, and (d) *K* = 0.6. Parameters are *ε*=0.1, $C_l$ = 0.05, and Λ=4.

Additionally, $\dot{\gamma}_w \equiv \dot{\gamma}(Y = \pm 1, Z)$ is also shown in Figure 3c, along with $\dot{\gamma}_m \equiv \dot{\gamma}(Y = 0, Z)$ in Figure 3d, as functions of *Z*. As described above, the decrease in the magnitude of the rate-of-deformation tensor everywhere on the flow domain and the boundaries with increasing slip coefficient is clearly evident. Note that in contrast to the no-slip case, where a constant $\dot{\gamma}_m$ is established within a short distance from the inlet, the presence of slip leads to a smooth spatial variation of the strain rate; that is, the flow remains purely extensional in character, but the strain rate is no longer constant. In the no-slip case, $\dot{\gamma}_m \equiv \dot{\gamma}(Y=0,Z)$ is primarily determined mainly by the leading-order term, i.e. the quantity 3(Λ-1)/4 (Sialmas & Housiadas 2024; Housiadas *et al.* 2025) while the next two terms introduce a Z-dependence through the shape function *H=H(Z)*, which is more pronounced near the inlet.

In Figure 4, we present contour plots of $\dot{\gamma}(Y,Z)$ on the (*Y*,*Z*) plain. The influence of increasing slip coefficient on the magnitude of the rate-of-deformation tensor is shown for a fixed contraction ratio (Λ=4), aspect ratio (ε=0.1), and interaction coefficient ($C_I$=0.05). The cases correspond to no slip (*K* = 0, Fig. 4a), low slip (*K* = 0.15, Fig. 4b), medium slip (*K* =0.3, Fig. 4c), and strong slip (*K* = 0.6, Fig. 4c). As expected, the magnitude of $\dot{\gamma}$ increases near the walls and toward the exit of the channel, and a general increase is observed as *K* increases.

### 4.2. Initial conditions and solution in the entrance region

The solution in the entrance region is required as an initial condition for the equations in the hyperbolic section of the channel. Considering the fully developed flow as determined by Sialmas & Housiadas (2024) (Eq. 27), and assuming no spatial variations of the orientation tensor components, Eqs. (18)–(21) simplify substantially to algebraic equations, which can be solved analytically for the quadratic closure or numerically for the hybrid closure, as demonstrated by Housiadas *et al.* (2025). As found for the no-slip case, only the interaction coefficient affects the results, since the flow remains purely shear in nature, although with a lower shear rate due to wall slip, as given in Eq. (27). Indeed, one can easily show that for pure shear flow with an imposed shear rate $\dot{\gamma}$, the governing equations for the orientation tensor reduce to the same form with $\dot{\gamma}^{(e)}$ replacing $\dot{\gamma}$. Consequently, the solutions for pure shear flow and fully developed Poiseuille flow in a straight channel are identical.

Worth mentioning is that on the midplane (*Y* = 0) the magnitude of the rate-of-deformation tensor is zero, and therefore the solution appears to be undetermined at that location. However, the diagonal components of the orientation tensor can be obtained from their limiting values just above and below the midplane. For the off-diagonal component, as the two limiting values above and below the midplane differ, an option is to set it to zero (Housiadas *et al.* (2025)). This indicates that a discontinuity in one of the nontrivial diagonal components of the orientation tensor at a point on the inlet cannot, in any case, be avoided.

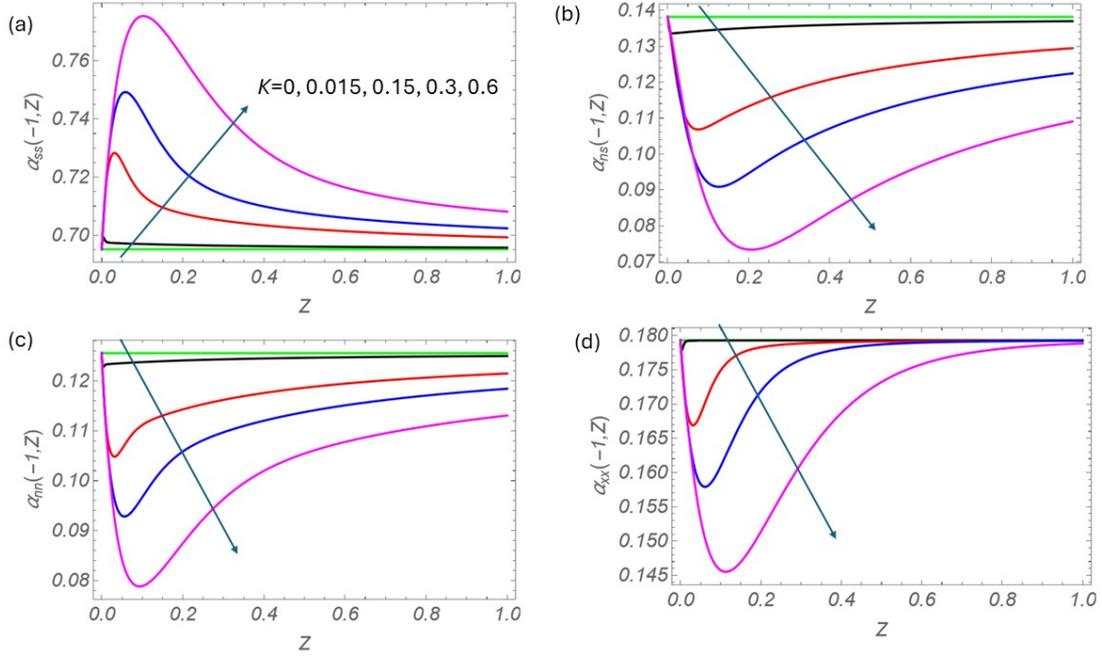

**Figure 5:** Components of the orientation tensor along the lower wall (*Y*=-1) as functions of the *Z* coordinate. Parameters are $\varepsilon$=0.1, $\Lambda$=4, and $C_I$ = 0.05. The arrow indicates the direction of increasing slip coefficient.

### 4.3. Solution at the walls of the hyperbolic region

Recalling that for any function $f = f(Y,Z)$, its convective derivative along the channel walls ($Y = \pm 1$) simplifies to $Df/Dt = U_w \, df/dZ$ where $U_w(Z) = H^{-1}(Z) \partial \Psi / \partial Y \big|_{Y=\pm 1}$ is the slip velocity. In this case, Eqs. (18)–(21) reduce to a nonlinear system of ODEs with unknowns $a_{ss}(Y_0,Z)$, $a_{ns}(Y_0,Z)$, and $a_{nn}(Y_0,Z)$ for $Y_0 = \pm 1$. When the velocity field is known, this system can be solved *a priori* to determine the evolution of the orientation tensor along the walls. As previously concluded for the no-slip case (*K* = 0, for which $U_w(Z) = 0$), the wall solution is fundamental and plays a critical role in the evolution of the orientation tensor within the varying section of the channel. The ODEs are therefore solved numerically, using as initial conditions at Z=0 the solution obtained in the entrance region—namely, the solution for fully developed Poiseuille flow (which, as reiterated, exactly matches that for pure shear flow). In this manner, the orientation tensor remains continuous throughout the entire channel, encompassing both the entrance and varying sections, while exhibiting jump discontinuities only in its Z-derivative at Z=0, due to the change in the slope of the shape function ($H'(0^-) = 0$, $H'(0^+) = -(\Lambda - 1)$).

The results of the calculations for the rotated components on the lower wall are illustrated in Figure 5 as a function of the distance from the inlet, $Z$; for the upper wall, the results are the same, but with the sign of the off-diagonal component reversed. As can be seen, the sudden contribution of the convective term in the equation for the orientation tensor—which is zero in the entrance region—results in an abrupt alignment of the fibers tangentially to the walls, thereby generating a boundary layer near the inlet. The larger the slip coefficient, the greater the slip velocity, and the stronger the alignment of the fibers. This phenomenon reaches a maximum within a short distance from the inlet and then decreases monotonically and smoothly as the fluid slips along the wall, approaching an orientation state similar to that at the inlet.

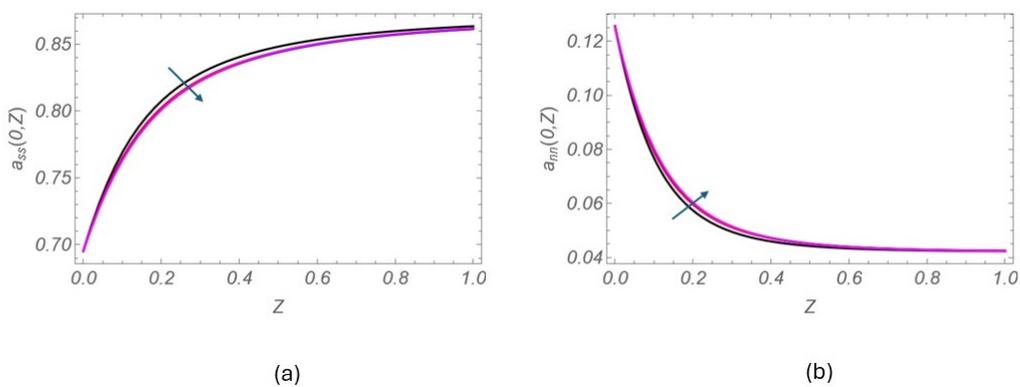

(a)                      (b)

**Figure 6:** The evolution of the diagonal components of the orientation tensor on the midplane as functions of the Z-coordinate. (a) $\alpha_{ss}$; (b) $\alpha_{nn}$. The off-diagonal component $\alpha_{ns}$ is zero due to its antisymmetry with respect to $Y=0$. The results are shown for $K=0$, 0.15, 0.3, and 0.6. The other parameters are $\varepsilon=0.1$, $\Lambda=4$, and $C_I=0.05$. Arrows indicate the direction of increasing $K$.

### 4.4. Solution in the midplane of the hyperbolic region

Eqs. (18)-(21) can also be solved *apriori* on the midplane of the channel ($Y=0$), independently of the remaining equations. Indeed, as for the channel walls, the convective derivative of any function $f = f(Y,Z)$ on the midplane simplifies to $Df/Dt = U_m\, df/dZ$ where $U_m(Z) = H^{-1}(Z) \partial \Psi / \partial Y \big|_{Y=0}$ is the midplane fluid velocity. Thus, Eqs. (18)-(21) reduce to a system of first-order ODEs along the axial direction with unknowns $a_{ss}(0,Z)$, $a_{ns}(0,Z)$, and $a_{nn}(0,Z)$. This system is subject to initial conditions at Z=0 which must be compatible with the corresponding simple Poiseuille solution at the entrance. As explained above, however, in the midplane of the entrance region all the components of the rate of deformation tensor become zero and thus the components of the orientation tensor become undetermined. The diagonal components of the orientation tensor which are symmetric with respect to the midplane, are taken to be their limiting values above and below the midplane, while the off-diagonal component $a_{ns}$, which is antisymmetric with respect to the midplane, is taken to be zero, i.e., $a_{ns}(Y=0, Z=0) = 0$. Taking into account the symmetry conditions at Y=0, the solution of the solution of Eq. (19) along the midplane is $a_{ns}(Y=0, Z) = 0$ for $0 \leq Z \leq 1$.

Results for the evolution of the components of the orientation tensor are represented in Figure 6 for slip coefficients $K$ = 0, 0.15, 0.3, and 0.6; the other parameters are $\varepsilon = 0.1$, Λ=4, and $C_I = 0.05$. One can see that the strong elongational character of the flow leads to a monotonic increase of $a_{ss}(0, Z)$ with Z; the increase is faster close to the inlet, i.e. as the fibers enter the hyperbolic section of the channel and are forced to align with the flow. In contrast, $a_{nn}(0, Z)$ decrease monotonically with the distance from the inlet. Also, as mentioned above, $a_{ns}(0, Z) = 0$ for any Z.

Regarding the effect of K, it is evident that it is negligible, since the curves for all slip cases are practically indistinguishable. This is somewhat unexpected, since—as shown in Figure 4d—increasing the slip coefficient substantially reduces the magnitude of the rate-of-deformation tensor $\dot{\gamma}_m \equiv \dot{\gamma}(Y=0, Z)$, which, on the midplane, coincides with the stretching rate in the main flow direction, $\dot{\gamma}(Y=0, Z) \equiv \partial U / \partial Z \big|_{Y=0}$. By examining the governing ODEs more closely, one can see that the key kinematic quantity determining the downstream evolution of the fiber orientation is the ratio $U_m / \dot{\gamma}_m$ (and not $\dot{\gamma}_m$ alone), which varies linearly with Z and shows negligible dependence on the slip coefficient.

### 4.5. Effect of the slip coefficient

For the standard no-slip case, Housiadas *et al.* (2025) investigated parametrically the effects of the channel's aspect ratio, $\varepsilon$, the contraction ratio, $\Lambda$, and the interaction coefficient, $C_I$, on the components of the orientation tensor. Here, we present and discuss the influence of the dimensionless slip coefficient, using the typical values $\varepsilon = 0.1$, $\Lambda = 4$, and $C_I = 0.05$. Figure 7 shows the simulation results at the outlet cross-section as functions of Y. Four cases are illustrated to demonstrate the influence of fluid slip along the walls, ranging from no slip ($K = 0$) to low slip ($K = 0.15$), moderate slip ($K = 0.30$), and high slip ($K = 0.6$). First, it is evident that all components vary monotonically from their values at the walls to their corresponding values at the midplane. This highlights the significance of the wall and midplane solutions, which, as discussed above, can be computed a priori without solving the governing equations over the entire flow domain. Additionally, the transition from the wall to midplane values are localized in a narrow strip which becomes larger, with increasing slip coefficient.

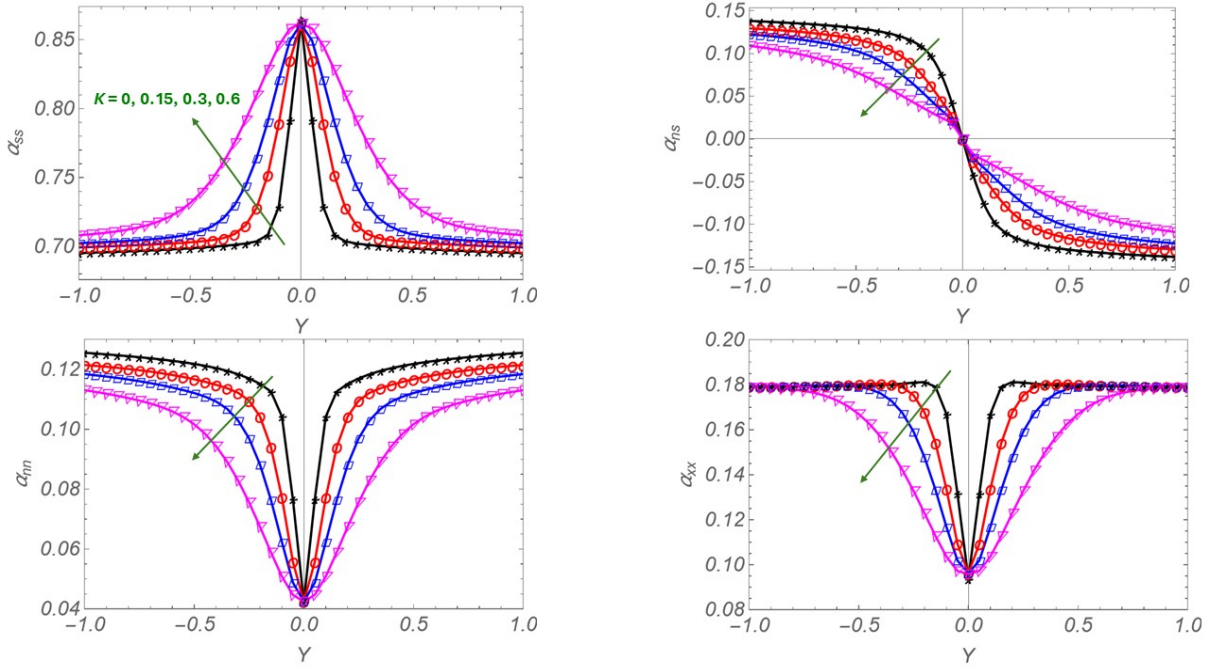

**Figure 7:** Effect of $K$ on the components of the orientation tensor as functions of $Y$ at the outlet ($Z=1$) of the hyperbolic section; (a) $\alpha_{ss}$, (b) $\alpha_{ns}$, (c) $\alpha_{nn}$, and (d) $\alpha_{xx}$. Parameters are $C_l=0.05$, $\Lambda=4$, and $\varepsilon=0.1$. The arrow shows the direction of increasing $K$.

We also observe in Figure 7 that, in contrast to the no-slip case, where the orientation tensor along the walls is independent of the magnitude of the rate-of-deformation tensor $\dot{\gamma}$, in the slip case the orientation is substantially affected by $\dot{\gamma}$ due to its strong dependence on the slip coefficient (as clearly seen in Figs 3 and 4). However, in the extensionally dominated region of the flow—i.e., along the midplane and its vicinity, where the orientation tensor is strongly influenced by the strain rate—the effect of the slip coefficient is imperceptible. Very similar profiles for the major diagonal component $\alpha_{ss}$ have been observed in the full 3D transient numerical simulations of injection molding of short-fiber composites by Jiang et al. (2022); note that the latter authors do not report the other components.

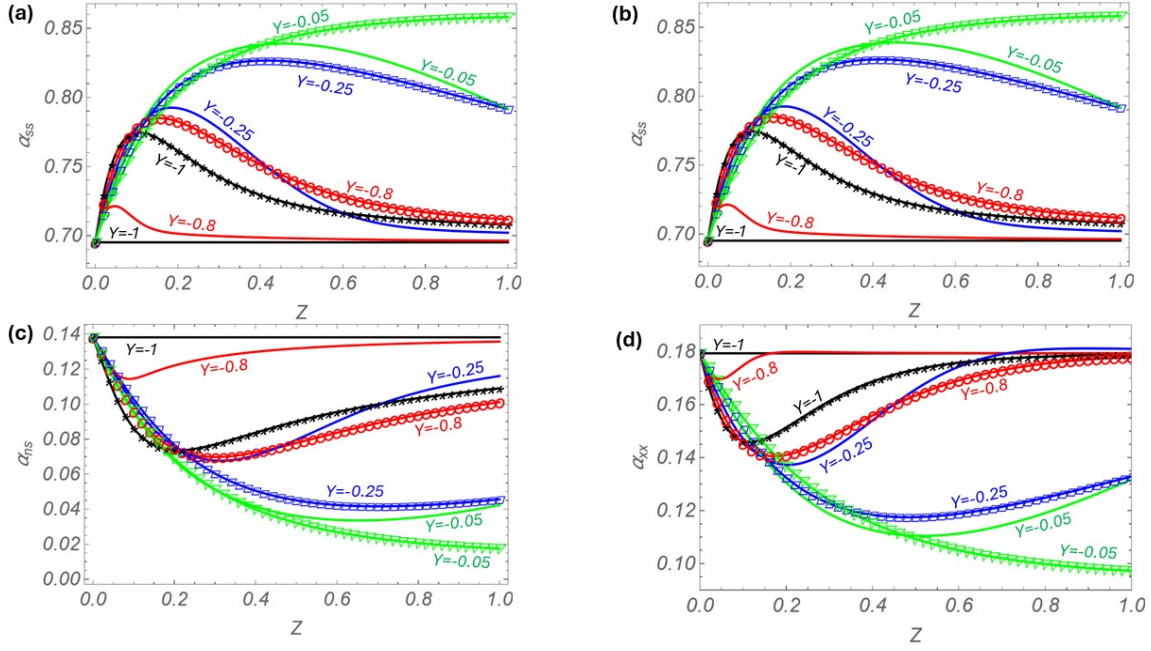

**Figure 8:** Evolution of the components of the orientation tensor as function of the distance from the inlet, Z, at various Y-locations. Parameters are: $C_I$ =0.05, $\varepsilon$=0.1, and Λ=4. The solid lines correspond to the no-slip case ($K = 0$), whereas the lines with symbols correspond to the strong slip case ($K = 0.6$).

### 4.6. Spatial evolution of the orientation tensor

Figure 8 presents the variations of the components of the orientation tensor along the main flow direction at various values of the mapped Y-coordinate, from Y=−1 (bottom wall) up to Y=-0.05 (almost at the midplane). The parameters are the same as in Figures 5-7. All components of the orientation tensor vary as functions of the distance from the inlet, always remaining between their corresponding constant values at the wall(s) and the values in the midplane. The results also reveal that the closer the fibers to the wall, the closer the corresponding values to that for pure shear flow. In fact, all curves increase, reaching a maximum at a short distance from the inlet of the hyperbolic section of the channel and then decrease monotonically approaching the solution at the wall, namely the solution which corresponds to the solution along the wall(s). However, in the midplane, where the flow is purely extensional in character, all diagonal components of the orientation tensor vary monotonically with Z, as shown previously in Figure 7. Additional information, in the form of contour plots, about the primary diagonal and off-diagonal components of the orientation tensor over the entire flow domain is provided in Figure 9. Specifically, the fibers become increasingly aligned with the main flow direction, as more clearly illustrated by the angle they form with respect to the midplane. The angle can be calculated by finding the angle between the normalized eigenvector corresponding to the major eigenvalue of the orientation tensor and the unit vector along the midplane, $\mathbf{e}_z$, i.e. $\varphi = \cos^{-1}(\mathbf{v}_1 \cdot \mathbf{e}_z)$; see below in Figure 9.

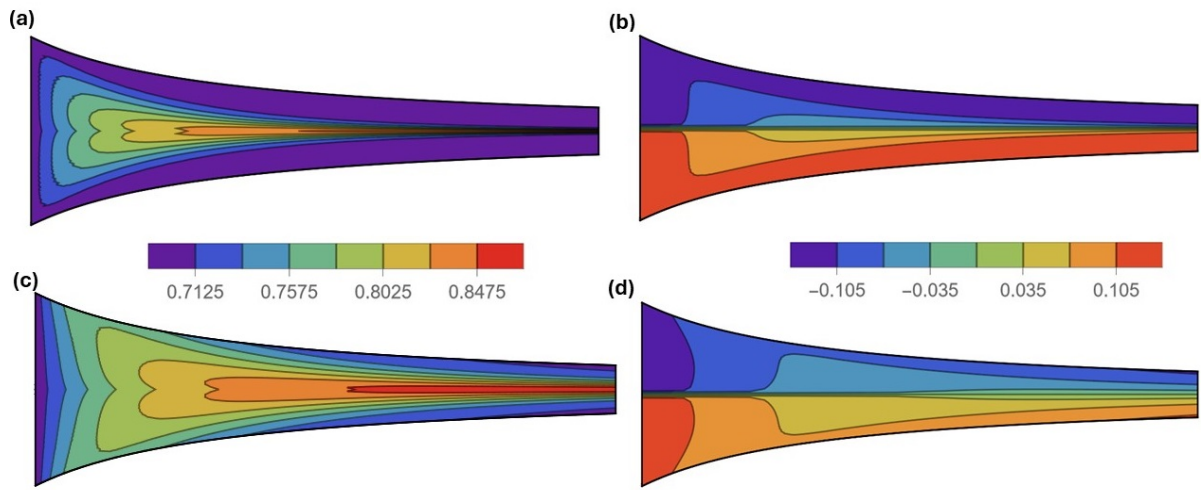

**Figure 9:** Contour plots of the primary diagonal component of the orientation tensor, $\alpha_{ss}$ (a, c), and the off-diagonal component, $\alpha_{ns}$ (b, d). Parameters: $C_I$ = 0.05, $\varepsilon$ = 0.1, and $\Lambda$ = 4.
Top panels: no-slip case ($K$ = 0); bottom panels: slip case ($K$ = 0.6).
Figures (a) and (c) share the same legend, as do Figures (b) and (d).

### *4.7. Eigenvector analysis of the orientation tensor*

Figure 10 shows the angle (in degrees) between $\mathbf{v}_1$ and $\mathbf{e}_z$, $\varphi = \cos^{-1}(\mathbf{v}_1 \cdot \mathbf{e}_z)$, as function of *Y* at the outlet of the channel, where $\mathbf{v}_1$ is the normalized eigenvector which corresponds to the largest eigenvalue of $\mathbf{a}$. In the same figure, the largest eigenvalue is also shown on the right vertical axis. Note that the angle is positive below the midplane and negative above the midplane, due to its antisymmetry with respect to *Y*=0, i.e., $\varphi(-Y) = -\varphi(Y)$. The results, which are presented for K = 0, 0.15, 0.30, and 0.60, show that the maximum angle appears at the wall and decreases monotonically with the distance from the midplane. Close to the midplane the angle approaches zero, i.e., $\mathbf{v}_1$ becomes parallel to $\mathbf{e}_z$, and the fibers are aligned with the main flow direction. Regarding the effect of the slip coefficient, increasing its magnitude has a pronounced influence on both the orientation angle and the major eigenvalue. As clearly seen, larger slip coefficients lead to smaller orientation angles, and larger values of the main eigenvalue. This demonstrates that wall slip combined with the channel hyperbolic geometry promotes faster and stronger alignment of the fibers with the main flow direction. In the case of pronounced slip, the variation from the wall to the midplane values becomes monotonic.

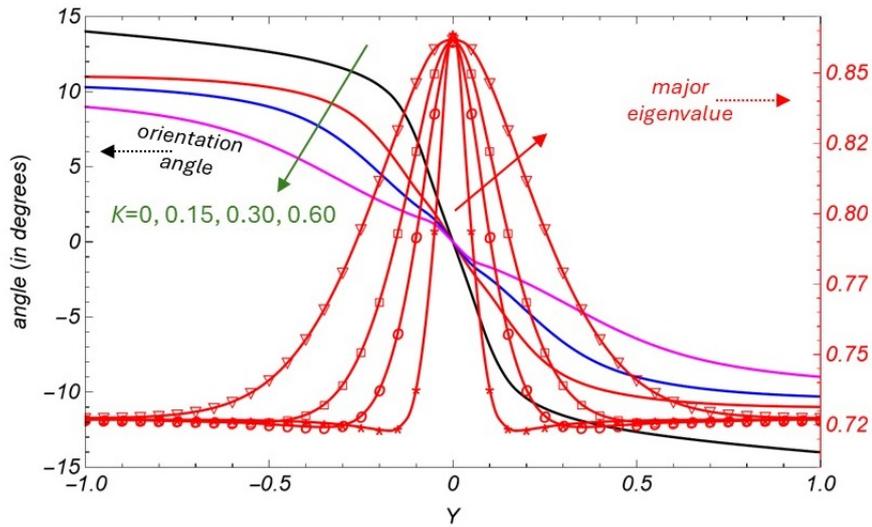

**Figure 10:** Left axis: the angle of the major eigenvector of the orientation tensor with respect to the midplane (solid lines). Right axis: the major eigenvalue (solid red lines with symbols). The results are shown at the exit of the channel ($Z=1$) for $\varepsilon=0.1$, $C_i=0.05$, and $\Lambda=4$. The solid arrows show the direction of increasing slip coefficient ($K = 0, 0.15, 0.30, 0.60$).

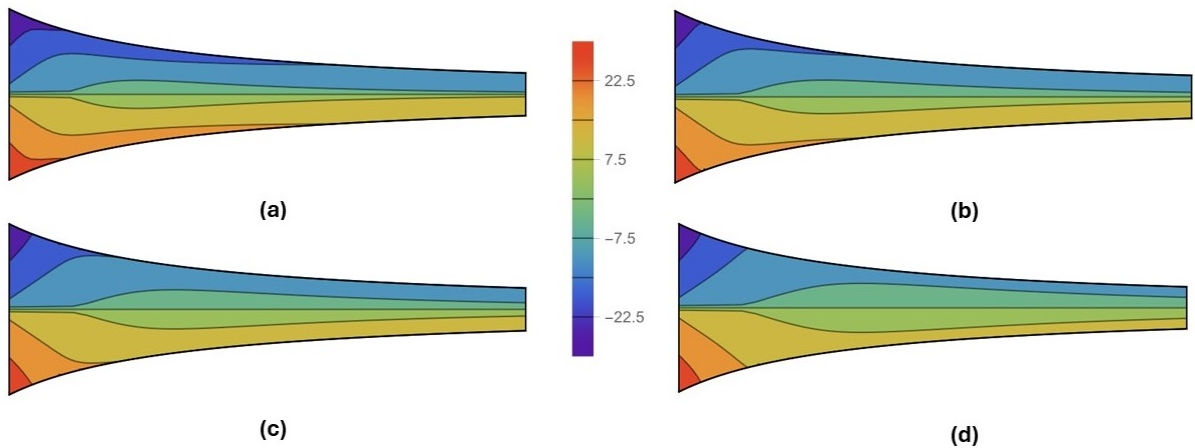

**Figure 11:** Contour plots of the angle (in degrees) of the major eigenvector of the orientation tensor with respect to the midplane. Parameters $\varepsilon=0.1$, $C_i=0.05$, $\Lambda=4$.
(a) $K = 0$; (b) $K = 0.15$; (c) $K = 0.30$; (d) $K = 0.6$.

Finally, contour plots of the orientation angle over the entire flow domain are depicted in Figure 11, providing an overview across the channel. One can see that large values of the orientation angle (in magnitude) are predicted near the entrance of the channel; however, the angle decreases continuously as the fibers are carried by the matrix fluid toward the channel exit, resulting in more fibers becoming aligned with the main flow direction.

## 5. Conclusions

In this work, we investigated the effect of wall slip on the creeping flow of a short-fiber suspension in a Newtonian fluid through a hyperbolic planar channel, employing a semi-analytical–numerical procedure. At the midplane, as in the corresponding no-slip case examined in our previous study, the flow remains purely extensional in character, and the fibers exhibit a pronounced orientation along the flow direction. Interestingly, although the extensional deformation rate on the midplane decreases significantly due to wall slip, the degree of fiber orientation remains practically unchanged. In contrast, wall slip exerts a strong influence on the fiber orientation in the rest of the flow domain, with its effect increasing with the slip coefficient. Specifically, the region near the midplane—where the orientation approaches that evaluated for the purely extensional flow—expands markedly with increasing slip, accompanied by a simultaneous overall increase in the degree of fiber alignment throughout the flow field.

The slip effects identified in this study complement those induced by variations in the other parameters governing hyperbolic short-fiber suspension flows, namely the two geometric parameters (the contraction and aspect ratios) and the fiber interaction parameter. Consequently, wall slip provides an additional mechanism to tailor fiber orientation in the final extruded product.

The present analysis thus represents a step toward a more comprehensive and physically faithful simulation framework for short-fiber composite processing, which will ultimately need to incorporate further effects such as the viscoelasticity of the polymer matrix, the extra-stress contribution of fibers to the total stress, and the influence of variable temperature in non-isothermal applications.

**ACKNOWLEDGMENTS**
Research was sponsored by the Army Research Laboratory and was accomplished under Cooperative Agreement Number W911NF-24-2-0127. The views and conclusions contained in this document are those of the authors and should not be interpreted as representing the official policies, either expressed or implied, of the Army Research Laboratory or the U.S. Government. The U.S. Government is authorized to reproduce and distribute reprints for Government purposes notwithstanding any copyright notation here.

**APPENDIX A**

In Eq. (18), the stretching functions $S_s^{(ss)}$ and $S_s^{(ss)}$, and $(\mathbf{A}^{(L)} : \dot{\boldsymbol{\gamma}})_{ss}$ are given as follows:

$$S_s^{(ss)} = \beta\varepsilon\frac{1-c^2}{1+c^2}\frac{\partial \Psi}{\partial Y} + \frac{\varepsilon}{H}\frac{\partial^2 \Psi}{\partial Y \partial Z} + \frac{c\varepsilon^2}{1+c^2}\frac{\partial^2 \Psi}{\partial Z^2} \tag{A1}$$

$$S_n^{(ss)} = \frac{4c\beta\varepsilon}{1+c^2}\frac{\partial \Psi}{\partial Y} + \left(cY\beta\varepsilon + \frac{1}{H^2}\right)\frac{\partial^2 \Psi}{\partial Y^2} + \frac{\beta\varepsilon^2 H}{1+c^2}\frac{\partial \Psi}{\partial Z} + 2Y\beta\varepsilon^2 H\frac{\partial^2 \Psi}{\partial Y \partial Z} + \frac{c^2\varepsilon^2}{1+c^2}\frac{\partial^2 \Psi}{\partial Z^2} \tag{A2}$$

$$(\mathbf{A}^{(L)} : \dot{\boldsymbol{\gamma}})_{ss} = \frac{1}{7}\left(a_{nn} + a_{ss} - \frac{1}{5}\right)\dot{\gamma}_{yy} + \frac{6}{7}a_{ns}\dot{\gamma}_{yz} + \frac{3}{7}\left(2a_{ss} - \frac{1}{5}\right)\dot{\gamma}_{zz} \tag{A3}$$

Similarly, in Eq. (19), $S_s^{(ns)}$ and $S_s^{(ns)}$, and $(\mathbf{A}^{(L)} : \dot{\boldsymbol{\gamma}})_{ns}$ are given as follows:

$$S_s^{(ns)} = \frac{\beta\varepsilon^2 H}{1+c^2}\frac{\partial \Psi}{\partial Z} + \frac{\varepsilon^2}{1+c^2}\frac{\partial^2 \Psi}{\partial Z^2} \tag{A4}$$

$$S_n^{(ns)} = \frac{4c\beta\varepsilon}{1+c^2}\frac{\partial \Psi}{\partial Y} + \left(cY\beta\varepsilon + \frac{1}{H^2}\right)\frac{\partial^2 \Psi}{\partial Y^2} + \frac{\beta\varepsilon^2 H}{1+c^2}\frac{\partial \Psi}{\partial Z} + 2Y\beta\varepsilon^2 H\frac{\partial^2 \Psi}{\partial Y \partial Z} + \frac{c^2\varepsilon^2}{1+c^2}\frac{\partial^2 \Psi}{\partial Z^2} \tag{A5}$$

$$(\mathbf{A}^{(L)} : \dot{\boldsymbol{\gamma}})_{ns} = \frac{3}{7}a_{ns}\dot{\gamma}_{yy} + \frac{2}{7}\left(a_{nn} + a_{ss} - \frac{1}{5}\right)\dot{\gamma}_{yz} + \frac{3}{7}a_{ns}\dot{\gamma}_{zz} \tag{A6}$$

Likewise, in Eq. (20), $S_s^{(nn)}$ and $S_s^{(nn)}$, and $(\mathbf{A}^{(L)} : \dot{\boldsymbol{\gamma}})_{nn}$ are given as follows:

$$S_s^{(nn)} = \frac{\beta\varepsilon^2 H}{1+c^2}\frac{\partial \Psi}{\partial Z} + \frac{\varepsilon^2}{1+c^2}\frac{\partial^2 \Psi}{\partial Z^2} \tag{A7}$$

$$S_n^{(nn)} = \frac{1-c^2}{1+c^2}\beta\varepsilon\frac{\partial \Psi}{\partial Y} + \frac{\varepsilon}{H}\frac{\partial^2 \Psi}{\partial Y \partial Z} + \frac{c\varepsilon^2}{1+c^2}\frac{\partial^2 \Psi}{\partial Z^2} \tag{A8}$$

$$(\mathbf{A}^{(L)} : \dot{\boldsymbol{\gamma}})_{nn} = \frac{3}{7}\left(2a_{nn} - \frac{1}{5}\right)\dot{\gamma}_{yy} + \frac{6}{7}a_{ns}\dot{\gamma}_{yz} + \frac{1}{7}\left(a_{nn} + a_{ss} - \frac{1}{5}\right)\dot{\gamma}_{zz} \tag{A9}$$

Last, in Eq. (21), $(\mathbf{A}^{(L)} : \dot{\boldsymbol{\gamma}})_{xx}$ is given as follows:

$$(\mathbf{A}^{(L)} : \dot{\boldsymbol{\gamma}})_{xx} = \frac{1}{7}\left(a_{xx} + a_{nn} - \frac{1}{5}\right)\dot{\gamma}_{yy} + \frac{2}{7}a_{ns}\dot{\gamma}_{yz} + \frac{1}{7}\left(a_{xx} + a_{ss} - \frac{1}{5}\right)\dot{\gamma}_{zz} \tag{A10}$$

Finally, the $O(\varepsilon^2)$ and $O(\varepsilon^4)$ terms for the streamfunction given in Eq. (24) are:

$$\psi_2 = \frac{3\left(2H^4 + 15H^3 K + 78H^2 K^2 + 297H K^3 + 360 K^4\right)}{40(H+3K)^4} - \frac{3H\left(H^2 + 3HK + 3K^2\right)Y^2}{20(H+3K)^3}$$

and $\psi_4 = \psi_{4,0} + Y^2 \psi_{4,2} + Y^4 \psi_{4,4}$, where:

$$\psi_{4,0} = \frac{3(112H^7 + 3775H^6 K + 31867H^5 K^2 + 62298H^4 K^3 - 350910H^3 K^4 - 1685817H^2 K^5 - 1811565H K^6 - 272160K^7)}{5600(H+3K)^7}$$

$$\psi_{4,2} = -\frac{3\left(28H^6 + 534H^5 K + 3870H^4 K^2 + 11277H^3 K^3 + 8316H^2 K^4 + 14175H K^5 + 22680K^6\right)}{1400(H+3K)^6}$$

$$\psi_{4,4} = -\frac{9H^4 K}{140(H+3K)^5}$$

**APPENDIX B**

The dimensionless equation that can be used to study finite fiber aspect ratio effects is:

$$\frac{\delta \mathbf{a}}{\delta t} + \frac{\xi}{2}(\dot{\boldsymbol{\gamma}} \cdot \mathbf{a} + \mathbf{a} \cdot \dot{\boldsymbol{\gamma}}) + (1-\xi)\mathbf{A}:\dot{\boldsymbol{\gamma}} = 2C_I\,(\mathbf{I} - 3\mathbf{a})\dot{\gamma} \qquad (B1)$$

where $\xi = 1 - \lambda$ and $\lambda = (r^2 - 1)/(r^2 + 1)$ is the fiber shape factor given in terms of its aspect ratio, $r$. For high aspect ratio fibers, $0 \leq \xi \ll 1$, whereas $\xi = 0$ corresponds to $r \to \infty$. In this limit, Eq. (B1) reduces to the equation solved in the current paper. Short fiber suspensions used in relevant applications have large aspect ratios, typically $r > 20$. For such values, the particle shape parameter satisfies $0.995 < \lambda \leq 1$, i.e. $\xi < 0.005$. To quantify the influence of finite aspect ratio explicitly, Eq. (B1) is solved for finite values of $r$.

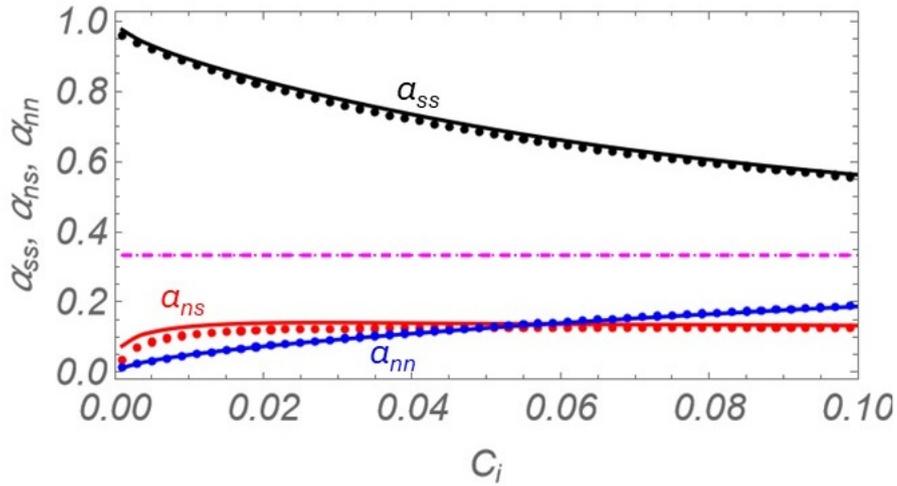

**Figure 12**: Orientation tensor components in the entrance region as a function of the interaction coefficient. Solid lines denote the fiber infinite aspect ratio case, circles correspond to $r$=10, and the horizontal line indicates 1/3 (the fully isotropic case).

First, the orientation tensor is calculated in the straight entrance region to determine the inlet conditions for the hyperbolic contraction. In this region, the flow is simple shear, the non-trivial components of Eq. (B1) reduce to algebraic equations, and the orientation solution is independent of the velocity field [Housiadas *et al.* (2025)]. Results are shown for the orientation-tensor components as functions of the interaction coefficient in the low-$C_I$ regime ($C_I < 0.1$). Differences between finite and infinite fiber aspect ratios appear only for small aspect ratio ($r$=10) and very low $C_I$ ($C_I < 0.04$), as shown in Figure 12; for $r \geq 20$ the solutions coincide with the fiber infinite–aspect-ratio results. These trends are in qualitative agreement with experimental and theoretical studies of rotary diffusion in simple shear [Folgar & Tucker (1984); Stover et al. (1992); Rahnama et al. (1995); Petrich et al. (2000)], which indicate that finite-aspect-ratio effects can be comparable to or larger than the rotary diffusion term. Nevertheless, for practical aspect ratios ($r \geq 20$), the infinite–aspect-ratio approximation Eq. (4) accurately captures fiber orientation trends.

Full contraction-flow simulations were performed for $r=10$ using the typical parameters $\varepsilon$ = 0.1, $K$ = 0.15, $C_I$ = 0.05, and $\Lambda$ = 4. The orientation-tensor components at the channel exit are shown in Figure 13. Deviations from the infinite–aspect-ratio results are minimal and are confined primarily to the $\alpha_{ss}$ component within the shear-dominated regions of the cross-section. Also, no qualitative differences in the spatial evolution of the orientation tensor are observed. Finally, simulations with higher fiber aspect ratios (of order twenty) yielded results that are indistinguishable from the corresponding infinite–aspect–ratio predictions.

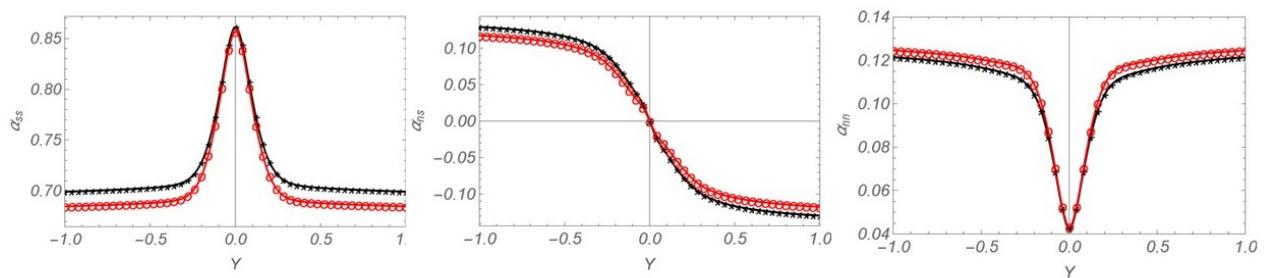

**Figure 13**: Orientation tensor components at the exit of the channel for finite and infinite fiber aspect ratios as a function of the mapped *Y*-coordinate. Star (black) symbols denote the infinite aspect ratio case; circles (red) correspond to $r = 10$. The other parameters are $\varepsilon$=0.1, $C_I$=0.05, $\Lambda$=4, $K$=0.15.